\documentstyle[aps,prb,epsf, multicol]{revtex}
\tighten
\begin{document}
\draft
\newcommand{\bn}{{\bf n}}
\newcommand{\bp}{{\bf p}}
\newcommand{\br}{{\bf r}}
\newcommand{\bq}{{\bf q}}
\newcommand{\bj}{{\bf j}}
\newcommand{\bE}{{\bf E}}
\newcommand{\eps}{\varepsilon}
\newcommand{\la}{\langle}
\newcommand{\ra}{\rangle}
\newcommand{\cK}{{\cal K}}
\newcommand{\cD}{{\cal D}}
\newcommand{\mybeginwide}{
    \end{multicols}\widetext
    \vspace*{-0.2truein}\noindent
    \hrulefill\hspace*{3.6truein}
}
\newcommand{\myendwide}{
    \hspace*{3.6truein}\noindent\hrulefill
    \begin{multicols}{2}\narrowtext\noindent
}

\title{
  Semiclassical theory of shot noise in disordered SN contacts 
}
\author{K.\ E.\ Nagaev$^{1,2}$ and M.\ B\"uttiker$^1$
}
\address{
  $^1$D\'epartement de Physique Th\'eorique, Universit\'e de Gen\`eve,
  CH-1211 Gen\`eve 4, Switzerland\\ 
  $^2$ Institute of Radioengineering and Electronics,
  Russian Academy of Sciences, Mokhovaya ulica 11, 103907 Moscow,
  Russia\\}
\date\today
\maketitle
\bigskip
\begin{abstract}

We present a semiclassical theory of shot noise in diffusive superconductor
- normal metal contacts.  At subgap voltages, we reproduce the doubling of
shot noise with respect to conventional normal-metal contacts, which is
interpreted in terms of an energy balance of electrons.  Above the gap, the
voltage dependence of the noise crosses over to the standard one with a
voltage-independent excess noise.  The semiclassical description of noise
leads to correlations between currents at different electrodes of
multiterminal SN contacts which are always of fermionic type, i.e.  negative.
Using a quantum extension of the Boltzmann - Langevin method, we reproduce
the peculiarity of noise at the Josephson frequency and obtain an
analytical frequency dependence of noise at above-gap voltages.

\end{abstract}

\begin{multicols}{2}
\narrowtext

In recent years, the noise properties of hybrid contacts involving
superconducting (S) and normal (N) metals attracted considerable
attention.\cite{Blanter-00}
Theoretical work in this direction was pioneered by
Khlus,\cite{Khlus} who found that Andreev reflection,\cite{Andreev} in
which electrons incoming from normal metals are reflected from 
the SN interface
as holes, play a key role in the shot noise of SN structures.  For clean SN
contacts, he predicted that the shot noise vanishes at subgap voltages
$eV < \Delta$.  de Jong and Beenakker\cite{deJong-94}
addressed dirty SN contacts and found that the shot noise at subgap
voltages is doubled with respect to its value in a normal contact with the
same resistance. This doubling of shot noise has 
already been experimentally confirmed.\cite{Jehl-99,Kozhevnikov-00}  
The distribution function of current fluctuations was
calculated and it was found that at subgap voltages, it describes
independent transfers of discrete charge $2e$ through the contact.\cite{Muzyka}
Lesovik {\it
et al.}\cite{Lesovik-99} found that the frequency dependence of the
shot noise exhibits a peculiarity at the Josephson frequency $\omega = 2eV$
instead of $\omega = eV$ in normal contacts.  More recently, the same
authors obtained additional singularities in the frequency dependence of the 
noise at above-gap voltages.\cite{Torres-00} These results could lead to 
the impression that the motion of electrons and holes is
correlated not only at the SN interface, but that the current is carried also 
through the
normal part of the contact in portions of $2e$ instead of $e$.  
Recent findings that the correlations of currents at
two normal contacts attached to a superconductor may be 
positive,\cite{Anantram-96,Martin-96} as in the
case of Bose statistics
seemed to reinforce such a view. On the other hand, 
from these works it is not clear whether the 
above results require phase coherence of electrons nor if 
there are restrictions for the contact length.

Recently, Gramespacher and one of the authors\cite{Gramespacher}
discussed the fluctuations in multiterminal SN contacts 
using electronlike and holelike distribution functions,
which were expressed in terms of 
quantum-mechanical injectivities\cite{Buttiker-93}
of the
terminals.  It was shown there, in particular, that positive correlations
vanish in the limit of a large channel number.  In this paper, we present a
semiclassical theory of shot noise in disordered SN contacts.  We show that
the doubling of shot noise may be obtained within a simple Boltzmann -
Langevin approach\cite{Kogan-69} provided that the appropriate boundary
conditions for the average distribution function are used.  We relate the
increase in the shot noise to the peculiar energy transport through an SN
interface and show that this effect is stable with respect to
phase-breaking. From the semiclassical nature of electron transport in the
N region, it immediately follows that the correlations between currents in
different electrodes of a multiterminal diffusive SN contact are always 
negative.  
We are
also able to explain the peculiarities of the frequency dependence of noise
in terms of the shape of the electron distribution function.


Consider a narrow normal-metal microbridge connecting massive normal and
superconducting electrodes.  The 
\begin{figure}
 \vspace{3mm}
 \centerline{
   \epsfxsize8cm
   \epsffile{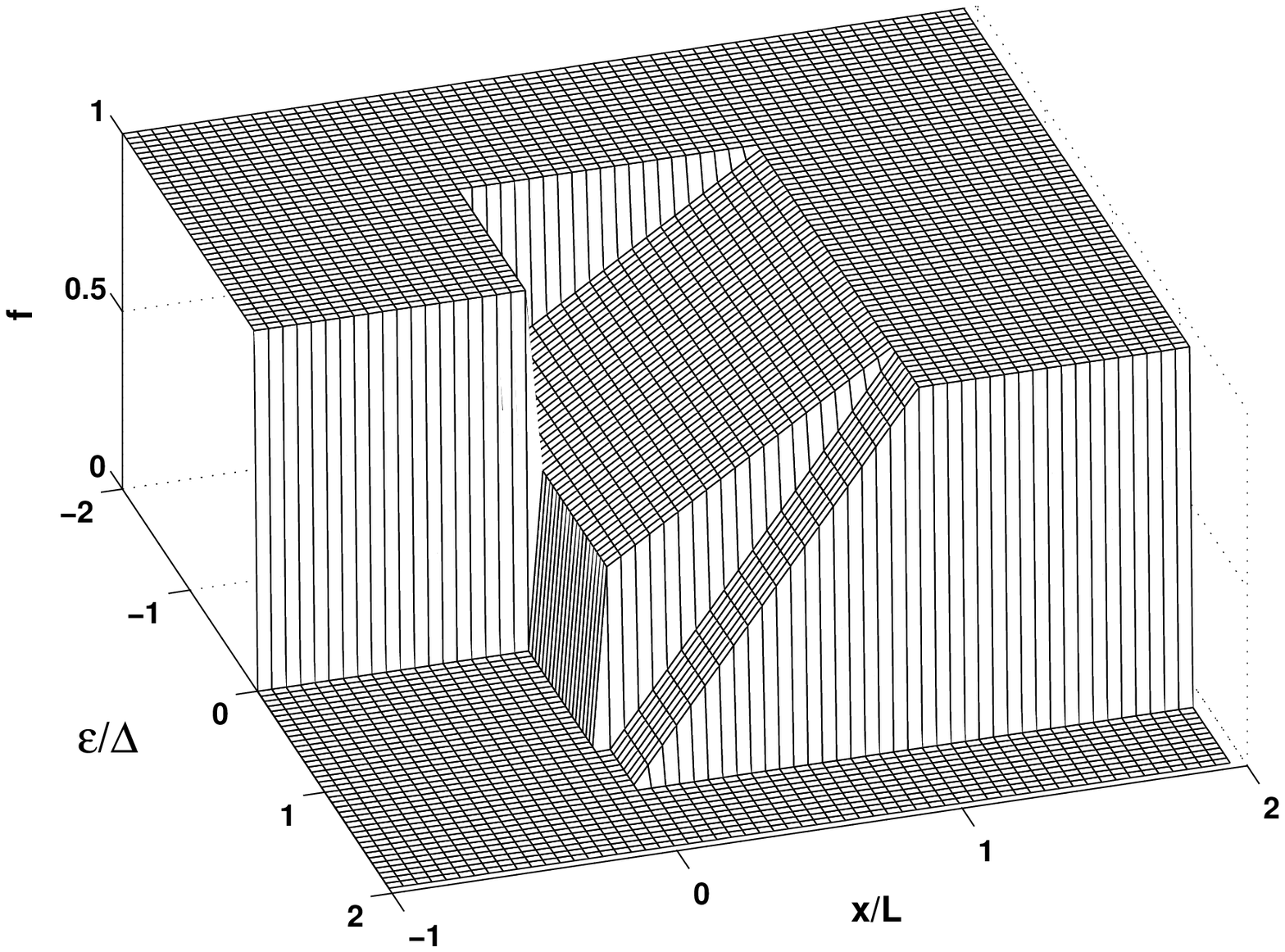}
 }
 \nopagebreak
 \caption{
   The non-equilibrium electron distribution function in the 
   normal part of an NS contact at voltage    
   $eV = 1.4\Delta$.
 }
\label{FIG.1}
\end{figure}
\noindent
elastic mean free path $l$ of electrons in
the microbridge is assumed to be short and the length of the contact $L$ is
assumed to be much larger than $(\hbar D/\Delta)^{1/2}$, where $\Delta$ is
the superconducting gap and $D$ is the diffusion coefficient in the normal
metal.  The applied voltage or temperature are considered to be much larger
than the Thouless energy $\eps_T = \hbar D/L^2$.  Under these conditions,
it is possible to neglect the penetration of the condensate into the
microbridge and consider it just as a normal metal with a nonequilibrium
distribution function of electrons.\cite{Artemenko-79,Nazarov-96}

Let the $x$ axis be directed along the contact, $x=0$ corresponding to the
superconducting electrode, and $x=L$, to the normal-metal one.  The
electron energy $\eps$ in the normal metal is measured from the Fermi level
of the superconductor.  Introduce an energy-dependent coefficient $W$ of
normal transmission of an electron from the contact through the NS
interface without being Andreev reflected, so that 
$W(\eps) = 0$ if $|\eps| < \Delta$ 
and $W(\eps) \to 1$,  for $|\eps| \gg \Delta $.  
 
Since the electrons in the superconducting electrode are described by an
equilibrium Fermi distribution function $f_0(\eps)$, the distribution 
function of electrons moving into the contact from the SN interface may be 
written in the form\cite{Artemenko-77,Blonder-82} 
\begin{eqnarray}
  f(\eps, v_x)
  =
  W f_0(\eps)
  +   
  (1 - W)
  [1 - f(-\eps, -v_x)],
\label{bound-1}
\end{eqnarray}
for $\eps > 0$ and $v_x > 0$, 
and the distribution function of holes moving into the contact from the 
interface may be written in the form 
\begin{eqnarray}
  1
  -
  f(\eps, v_x)
  =
  W[ 1 - f_0(\eps) ]
  +
  &&
  (1 - W)
  f(-\eps, -v_x),
\label{bound-2}
\end{eqnarray}
for $\eps < 0$ and $v_x > 0$. 
Since in the bulk of the contact the electron motion is diffusive, we must 
now obtain a suitable boundary condition at the NS interface for the diffusion 
equation. 

Introduce the symmetric and antisymmetric parts of $f$ with respect to $v_x$: 
\begin{eqnarray}
 f_s(\eps)
 = (1/2) 
 \left[
   f(\eps, v_x)
   +
   f(\eps, -v_x)
 \right],
\nonumber\\ 
 f_a(\eps)
 = (1/2) 
 \left[
   f(\eps, v_x)
   -
   f(\eps, -v_x)
 \right].
 \label{symm}
\end{eqnarray}
Inside the contact, the distribution function obeys the diffusion equation 
and may be represented as a sum of an isotropic part $f(\eps, x)$ and a small 
anisotropic part 
\begin{equation}
 f_1 
 =
 - (v_x/v_F)(l/L) 
 [
   f(\eps, L) - f(\eps, 0)
 ].
\label{f_1}
\end{equation}
Following Kupriyanov and Lukichev,\cite{Kupriyanov-88} we equate now 
$f(\eps,0)$ to $f_s(\eps)$ and $f_1$ to 
$(v_x/v_F)f_a(\eps)$. Introducing the distribution function of electrons in the 
normal electrode $f_n(\eps)$, one may express $f_a$ in terms of $f_s$ and 
$f_n$:
\begin{equation}
 f_a(\eps)
 =
 ({l}/{L})
 [
  f_s(\eps) - f_n(\eps)
 ].
\label{f_a}
\end{equation}

For the sake of definiteness, assume now $\eps > 0$. By substituting 
Eqs. 
(\ref{symm})  into Eqs. (\ref{bound-1}) for $f(\eps, v_x)$ and 
(\ref{bound-2}) 
for $1 - f(-\eps, v_x)$ and excluding $f_a$ from them by means of 
(\ref{f_a}),
one obtains a closed set of equations for $f_s(\eps)$ and $f_s(-\eps)$:
$$
 \left(
   1 + l/L
 \right)
 f_s(\eps)
 +
 ( 1 - W )
 \left(
   1 - l/L
 \right)
 f_s(-\eps)
 =
 1 - Wf_0(-\eps) 
$$
\begin{equation} 
 +
 (l/L)f_n(\eps)
 -
 (l/L)( 1 - W ) f_n(-\eps)
\label{f_s-2}
\end{equation}
and the equation obtained from it by interchanging $\eps$ and $-\eps$. The
solution of the set (\ref{f_s-2}) assumes different forms depending on the
relationship between $W$ and $l/L$.  In the subgap region $|\eps| <
\Delta$, $W = 0$ and we obtain
\begin{equation}
 f_s(\eps)
 =
 \frac{1}{2}
 \left[
   1 + f_n(\eps) - f_n(-\eps)
 \right]
\label{f_s-3}
\end{equation}
for both signs of $\eps$. Suppose that the normal electrode has a 
potential 
$V$. Then $f_n(\eps) = 
f_0(\eps -eV)$, and inside the gap,
\begin{equation}
 f_s(\eps)
 =
 \frac{1}{2}
 \left[
   f_0(\eps - eV) + f_0(\eps + eV)
 \right],
 \qquad
 |\eps| < \Delta .
\label{f_s-4}
\end{equation}
Outside the subgap region, $W \ne 0$, and $l/L$ may be neglected with 
respect to it virtually at all energies. In this case, one easily obtains 
\begin{equation}
 f_s(\eps) = f_0(\eps),
 \qquad
 |\eps| > \Delta . 
\label{f_s-5}
\end{equation}

Inside the contact, the distribution function is obtained by 
solving the diffusion equation with the boundary conditions (\ref{f_s-4}) 
and (\ref{f_s-5}). As a result, one obtains:
\mybeginwide
\begin{equation}
 f(\eps, x)
 =
 \left\{
   \begin{array}{ll}
    \displaystyle
    \frac{1}{2}
    \left(
      1 + \frac{x}{L}
    \right)
    f_0(\eps - eV)
    +
    \frac{1}{2} 
    \left(
      1 - \frac{x}{L}
    \right)
    f_0(\eps + eV),
    \qquad &
    |\eps| < \Delta
   \\   
    \displaystyle
    \rule{0pt}{10mm}
    \left(
      1 - \frac{x}{L}
    \right)
    f_0(\eps)
    +
    \frac{x}{L}
    f_0(\eps - eV),
    \qquad &
    |\eps| > \Delta.
   \end{array}
 \right.
 \label{f}
\end{equation}
\myendwide
At subgap voltages, the shape of the distribution function is exactly the same 
as if the contact were extended in negative direction from 0 to $-L$ and the 
voltage $-V$ were applied at this point.\cite{deJong-94} The distribution 
function for an above-gap voltage is shown in Fig. 1. Note that at finite 
voltages, $f(\eps)$ is discontinuous at $|\eps| = \Delta$ even at nonzero 
temperature.

    
Consider now the noise of the contact.  Since all the voltage drop takes
place inside the normal-metal microbridge and there is no voltage drop
across the SN interface, it is reasonable to assume that all the noise is
produced by random impurity scattering in the microbridge only.  In this
case, we can use the well known Langevin equations for the current
fluctuations,\cite{Nagaev-98}
\begin{equation}
 \delta\bj
 =
 -D
 \frac{ 
   \partial 
 }{ 
   \partial\br
 }
 \delta\rho
 -
 \sigma
 \frac{
   \partial
 }{
   \partial\br
 }
 \delta\phi
 +
 \delta\bj^{ext},
\label{dj}
\end{equation}
where $D$ is the diffusion coefficient, $\sigma$ is the electric
conductivity, $\delta\rho(\br)$ is the charge-density fluctuation,
$\delta\phi(\br)$ is the local fluctuation of the electric potential, and
the correlator of extraneous currents $\delta\bj^{ext}$ is given by
$$
 \la
  \delta j^{ext}_{\alpha}(\br_1)
  \delta j^{ext}_{\beta }(\br_2)
 \ra
 _{\omega}
 =
 4\sigma
 \delta_{\alpha\beta}
 \delta( \br_1 - \br_2 )
 T_N(\br_1),
$$
\begin{equation}
 T_N(\br)
 =
 \int d\eps\,
 f(\eps, \br)
 [
   1 - f(\eps, \br)
 ].
\label{T_N}
\end{equation}
At sufficiently low frequencies, the fluctuation of the total current
$\delta\bj$ may be considered as constant along the contact, and Eqn.
(\ref{dj}) may be integrated over its length.  As a result, the gradient
terms in (\ref{dj}) drop out because of the boundary conditions $\delta\rho
= 0$ and $\delta\phi =0$, and one arrives at the standard expression for
the noise in disordered metal contacts:\cite{Nagaev-92}
\begin{equation}
 S_I
 =
 \frac{4}{R}
 \int
 \frac{dx}{L}
 T_N(x),
\label{S_I}
\end{equation}
where $R$ is the normal resistance of the contact. 
Substituting now the distribution function (\ref{f}) into (\ref{S_I}) and 
performing the integration, one obtains that
\mybeginwide
$$
 S_I
 =
 4\frac{T}{R}
 \left\{
   \frac{2}{3}
   +
   \frac{1}{3}
   \frac{eV}{T}
   \coth
   \left(
     \frac{eV}{T}
   \right)
   +
   \frac{1}{6}
   \left[
     \tanh
     \left(
       \frac{\Delta + eV}{2T}
     \right)
     +
     \tanh
     \left(
       \frac{\Delta - eV}{2T}
     \right)
     -
     2\tanh
     \left(
       \frac{\Delta}{2T}
     \right)
   \right]
 \right.
$$
\begin{equation}
 \left.
   +
   \frac{1}{6}
   \left[
     \coth
     \left(
       \frac{eV}{2T}
     \right)
     -
     2\coth
     \left(
       \frac{eV}{T}
     \right)
   \right]
   \,
   \ln
   \left[
     \frac{
       \exp(\Delta/T) + \exp(eV/T)
     }{
       \exp(\Delta/T) + \exp(-eV/T)
     }
   \right]
 \right\}.
\label{S_I-2}
\end{equation}
\myendwide
At zero voltage, this expression reduces to the Nyquist formula $S_I = 4T/R$.
At zero temperature yet finite voltage, Eqn. (\ref{S_I-2}) takes a very simple 
form:
\begin{equation}
 S_I
 =
 \frac{4}{3}
 \frac{e|V|}{R}
 +
 \frac{2}{3}
 \theta( e|V| - \Delta )
 \frac{ \Delta - e|V| }{R}.
\label{S_I-3}
\end{equation}
The shot noise in diffusive SN contacts is doubled at 
voltages $e|V| < \Delta$ as compared to the same contacts between normal 
metals. The reasons for the shot-noise doubling can be understood in 
terms of the noise temperature of electrons $T_N$ and the energy transport. 
The electrons in the contact acquire additional energy because of their 
acceleration by the field. If both electrodes are made of normal metal, the 
excess energy flows into both electrodes thus effectively cooling the contact 
and decreasing $T_N$. However if one of the electrodes is superconducting, the 
SN boundary forbids heat transfer in the subgap region.\cite{Abrikosov} The 
absence of heat transfer through an SN boundary has been known for many years, 
and this is 
precisely what motivated the pioneering paper by Andreev.\cite{Andreev} Because 
there is only one energy drain now instead of two, naturally $T_N$ is higher 
than in the case of a normal contact. 

The above semiclassical derivation 
shows that the doubling of shot noise is insensitive to phase-breaking. In 
the case of strong electron-electron scattering, the effective-temperature 
approximation may be used at $eV \ll \Delta$ and precisely the same 
heat-balance equation for the effective temperature $T_e$ as in normal 
contacts\cite{Nagaev-95,Kozub-95} can be written except that 
the heat-absorption boundary condition $T_e = T$ at the SN interface should be 
replaced by the zero heat-flux condition $\partial T_e/\partial x = 0$. It is 
easily seen that this should double the shot noise making it  $2 \times 
(\sqrt{3}/2)eI = \sqrt{3}eI$ and thus increasing it above the 
noninteracting-electron value precisely as for a normal 
contact.\cite{Nagaev-95,Kozub-95} Much like in the case of normal contacts, the 
noise should be suppressed by energy relaxation caused e.\ g.\ by phonon 
emission.


Using the semiclassical approach, Sukhorukov and Loss\cite{Sukhorukov-99}  
recently investigated the correlations between currents in different 
contacts of multiterminal diffusive normal-metal systems. Here we extend their 
discussion to the case of multiterminal SN systems.\cite{Blanter-97} It is 
convenient to introduce characteristic potentials\cite{Buttiker-93} 
$\phi_n(\br)$, $n =1, \ldots, N$ associated whith each contact $n$ which equal 
1 at contact $n$ and are zero at all other electrodes obeying the equation
\begin{equation}
 \nabla
 (
   \sigma\nabla\phi_n
 )
 =
 0
\label{phi_n}
\end{equation}
in the bulk of the contact. In terms of these potentials, the normal-state 
conductance matrix of the contact may be written in the form
\begin{equation}
 G_{mn}
 =
 -\int d{\bf S}_m
 \cdot
 \sigma\phi_m\nabla\phi_n,
\label{G_mn}
\end{equation}
where $S_m$ is the interface of the system with electrode $m$,
and the cross-correlated spectral density for contacts $n$ and $m$ may be 
written in the form\cite{Sukhorukov-99} 
\begin{equation}
 S_{mn}
 =
 4
 \int d^3\br\,
 \nabla\phi_n
 \cdot
 \nabla\phi_m
 \cdot
 \sigma T_N(\br).
\label{S_mn}
\end{equation}
Performing twice an integration by parts and making use of (\ref{phi_n}) and 
a similar equation for $f$ in $T_N$, equation (\ref{S_mn}) is easily brought to 
a 
form
$$
 S_{mn}
 =
 -
 2G_{mn} 
 \left.
   T_N
 \right|
 _{S_m}
 -
 2G_{nm}
 \left.
   T_N
 \right|
 _{S_N}
$$
\begin{equation} 
 -
 4\int d^3 r\,
 \phi_n \phi_m \sigma
 \int d\eps\,
 (\nabla f)^2.
\label{S_mn-2}
\end{equation}
As $T_N$, $G_{mn}$, and $\phi_m$ are always positive, this proves that the 
current correlations are always negative, i.\ e.\ of fermionic type. As this 
proof holds for an arbitrary shape of the distribution function 
at the interfaces 
with electrodes, the correlations are negative for an arbitrary number of 
superconducting electrodes provided that the Josephson effect between them is 
suppressed.


Consider now the noise at frequencies $\omega \sim eV$. This is the range of 
quantum noise, which clearly cannot be described in terms of the Boltzmann 
equation. However it was recently shown that for disordered metals, the 
Langevin 
scheme may be extended to high frequencies by introducing a frequency-dependent 
noise temperature\cite{Nagaev-00} 
$$
 T_N(\br, \omega)
 =
 \frac{1}{2}
 \int d\eps
 \left\{
   f(\br, \eps)
   [
    1 - f(\br, \eps + \omega)
   ]
 \right.
$$
\begin{equation}
 \left.
   +
   f(\br, \eps + \omega)
   [
    1 - f(\br, \eps)
   ]
 \right\} 
\label{T_N-q}
\end{equation} 
into Eq. (\ref{T_N}). If the frequency is smaller than $1/RC$, where $C$ is 
the capacity of the contact to all possible external gates, the current 
fluctuation may still be considered as constant along the contact length, and 
we arrive again at Eq. (\ref{S_I}) with $T_N(x,\omega)$ substituted for 
$T_N(x)$. Note that the latter condition is well compatible with $\omega \gg 
D/L^2$ because $1/RC$ is larger by an 
extra factor containing the ratio of the contact diameter to the 
Thomas - Fermy screening length. 

Suppose first that $eV < \Delta$. Integration in (\ref{T_N-q}) and (\ref{S_I}) 
then gives
\begin{equation}
 S_I(\omega)
 = 
 \frac{1}{3R}
 [
  4 U(\omega)
  + U(2eV + \omega)
  + U(2eV - \omega)
  ],
\label{S_I-q1}
\end{equation}
with $U(\omega) = \omega\coth(\omega/2T)$, which reduces to
\begin{equation}
 S_I(\omega)
 = 
 ({1}/{3R})
 [
  4\omega
  +
  {\rm max}
  ( 4eV, 2\omega )
 ]
\label{S_I-q2}
\end{equation}
at zero temperature. The frequency dependence of $S_I$ exhibits a kink at 
$\omega = 2eV$, which would correspond to the Josephson frequency of the 
contact if the second electrode were also superconducting. The reason for the 
doubling of the kink frequency with respect to normal contacts is that the 
range of energies where the drop of the distrbution function takes place is two 
times larger now. 

In the case where $T=0$ and $|eV| > \Delta$, integration gives
$$
 S_I(\omega)
 =
 \frac{|\omega|}{R}
 +
 \frac{1}{6R}
 \left(
   |\omega + 2\Delta|
   +
   |\omega - 2\Delta|
   +
   |\omega - eV + \Delta| 
 \right.
$$
\begin{equation}
 \left. 
   +
   |\omega - eV - \Delta|  
   +
   |\omega + eV + \Delta|
   +
   |\omega + eV - \Delta|
 \right).
\label{S_I-q3}
\end{equation}
in accordance with recent numerical results,\cite{Torres-00} this expression 
exhibits peculiarities at $\omega = 2\Delta$ and $\omega = eV \pm \Delta$, 
which are related with discontinuities in the energy dependence of the 
distribution function. The slope of $S(\omega)$ increases from $1/R$ at $\omega 
\to 0$ to $2/R$ at $\omega \to \infty$.

In this work we have developped a semi-classical approach for the noise
in metallic diffusive NS-structures. This approach explains a number of key 
results in a physically transparent manner and demonstrates that these
results are immune to dephasing. 

This work was supported by the Swiss National Science Foundation.

\end{multicols}

\begin{thebibliography}{99}
%
\bibitem{Blanter-00} For a review see Y. Blanter and M. B\"uttiker, 
Phys. Rep. (unpublished). cond-mat/9910158
%
\bibitem{Khlus}
V. A. Khlus, Zh. Eksp Teor. Fiz. {\bf 93}, 2179 (1987) [Sov. Phys. JETP {\bf 
66}, 1243 (1987)].
%
\bibitem{Andreev}
A. F. Andreev, Zh. Eksp. Teor. Fiz. {\bf 46}, 1823 (1964) [Sov. Phys. JETP {\bf 
19}, 1228 (1964)].
%
\bibitem{deJong-94}
M. J. M. de Jong and C. W. J. Beenakker, Phys. Rev. B {\bf 49}, 16070 (1994).
%
\bibitem{Jehl-99}
X. Jehl, P. Payet-Burin, C. Baraduc, R. Calemczuk, and M. Sanguer, Phys. Rev. 
Lett. {\bf 83}, 1660 (1999).
%
\bibitem{Kozhevnikov-00}
A. A. Kozhevnikov, R. J. Shoelkopf and D. E. Prober, Phys. Rev. Lett. {\bf 84}, 
3398 (2000).
%
\bibitem{Muzyka}
B.A. Muzykantsky and D. E. Khmelnitskii, Phys. Rev. B {\bf 50}, 3982 (1994).
%
\bibitem{Lesovik-99}
G. B. Lesovik, T. Martin, and J. Torr\`es, Phys. Rev. B {\bf 60}, 11935 (1999).
%
\bibitem{Torres-00}
J. Torres, T. Martin, and G. B. Lesovik, cond-mat/0004489 (unpublished).
%
\bibitem{Anantram-96}
M. P. Anantram and S. Datta, Phys. Rev. B {\bf 53}, 16390 (1996).
%
\bibitem{Martin-96}
Th. Martin, Phys. Lett. A {\bf 220}, 137 (1996).
%
\bibitem{Gramespacher}
T. Gramespacher and M. B\"uttiker, Phys. Rev. B {\bf 61}, 8125 (2000).
%
\bibitem{Buttiker-93}
M. B\"uttiker, J. Phys.: Condens. Matter {\bf 5}, 9361 (1993).
%
\bibitem{Kogan-69}
Sh. M. Kogan and A. Ya. Shul'man, Zh. Eksp. Teor. Fiz. {\bf 56}, 862 (1969) 
[Sov. Phys. JETP {\bf 29}, 467 (1969)].
%
\bibitem{Artemenko-79}
S. N. Artemenko, A. F. Volkov, and A. V. Zaitsev, Solid State Commun. {\bf 30}, 
771 (1979).
%
\bibitem{Nazarov-96}
Yu. V. Nazarov and T. H. Stoof, Phys. Rev. Lett. {\bf 76}, 823 (1996).
%
\bibitem{Artemenko-77}
S. N. Artemenko and A. F. Volkov, Zh. Eksp. Teor. Fiz. {\bf 72}, 1018 (1977) 
[Sov. Phys. JETP {\bf 45}, 533 (1977)].
%
\bibitem{Blonder-82}
G. E. Blonder, M. Tinkham, and T. M. Klapwijk, Phys. Rev. B {\bf 25}, 4515 
(1982).
%
\bibitem{Kupriyanov-88}
M. Yu. Kupriyanov and V. F. Lukichev, Zh. Eksp. Teor. Fiz. {\bf 94}, 139 (1988) 
[Sov. Phys. JETP {\bf 67}, 1163 (1988)].
%
\bibitem{Nagaev-98}
K. E. Nagaev, Phys. Rev. B {\bf 57}, 4628 (1998).
%
\bibitem{Nagaev-92}
K. E. Nagaev, Phys. Lett. A {\bf 169}, 103 (1992).
%
\bibitem{Abrikosov}
A. A. Abrikosov, {\it Fundamentals of the Theory of Metals}, North Holland, 
Amsterdam, 1988.
%
\bibitem{Nagaev-95}
K. E. Nagaev, Phys. Rev. B {\bf 52}, 4740 (1995).
%
\bibitem{Kozub-95}
V. I. Kozub and A. M. Rudin, Phys. Rev. B {\bf 52}, 7853 (1995).
%
\bibitem{Sukhorukov-99}
E. V. Sukhorukov and D. Loss, Phys. Rev. B {\bf 59}, 13054 (1999).
%
\bibitem{Blanter-97}
For a quantum-mechanical derivation, see Ya. M. Blanter and M. B\"uttiker, 
Phys. Rev. B {\bf 55}, 2127 (1997).
%
\bibitem{Nagaev-00}
K. E. Nagaev, cond-mat/9908193, to appear in Phys. Rev. B {\bf 62} (2000).

\end{thebibliography}
\end{document}